\documentstyle[12pt]{article}

\makeatletter
\@addtoreset{equation}{section}
\makeatother

\let\l=\lambda     \let\r=\rho
\let\s=\sigma     
    \let\D=\Delta  
  \let\S=\Sigma   
 
\let\la=\label  
  
\def\nn{\nonumber} \def\bd{\begin{document}} \def\ed{\end{document}}
\def\ds{\documentstyle} \let\fr=\frac \let\bl=\bigl \let\br=\bigr
\let\Br=\Bigr \let\Bl=\Bigl
\let\bm=\bibitem
\let\na=\nabla
\let\pa=\partial \let\ov=\overline
\newcommand{\be}{\begin{equation}}
\newcommand{\ee}{\end{equation}}
\def\ba{\begin{array}}
\def\ea{\end{array}}
\newcommand{\bea}{\begin{eqnarray}}
\newcommand{\eea}{\end{eqnarray}}
\newcommand{\ra}{\rightarrow}
\newcommand{\oh}{\frac{1}{2}}
\newcommand{\lra}{\longrightarrow}
\newcommand{\Lra}{\Leftrightarrow}
\newcommand{\ap}{\alpha^\prime}
\newcommand{\bp}{\tilde \beta^\prime}
\newcommand{\tr}{{\rm tr} }
\newcommand{\sg}{\sqrt {-g}}
\newcommand{\rn}{r_0}
\newcommand{\rpl}{r_p}
\newcommand{\hoch}[1]{$\, ^{#1}$}
\newcommand{\srt}{\sqrt{3}}
\newcommand{\oosrt}{\frac{1}{\sqrt{3}}}
\newcommand{\rmi}{r_-}
\newcommand{\Dp}{\Delta_{+}}
\newcommand{\Dm}{\Delta_{-}}
\newcommand{\Tr}{{\rm Tr} }
\newcommand{\NP}{Nucl. Phys. }
\newcommand{\tamphys}{\it Center for Theoretical Physics\\
Physics Department \\ Texas A \& M University
\\ College Station, Texas 77843}
\newcommand{\auth}{J. Rahmfeld\hoch{\dagger}}

\begin{document}
\hfill{}
\hfill{}

\hfill{CTP-TAMU-51/95}

\hfill{hep-th/9512089}

\vspace{24pt}

\begin{center}
{ \large {\bf Extremal Black Holes as Bound States}}

\vspace{48pt}

\auth

\vspace{10pt}

{\tamphys}

\vspace{72pt}

\underline{ABSTRACT}

\end{center}

We consider a simple static extremal multi-black hole solution with
constituents charged under different $U(1)$ fields. Each of the constituents
by itself is an extremal dilatonic black hole of coupling $a=\srt$. For a
special case with two electrically and two magnetically charged black holes
the multi-black hole solution interpolates between the familiar
$a=\sqrt{3},1,\frac{1}{\sqrt{3}}$ and $0$ solutions, depending on how many
black holes are placed at infinity. This proves the hypothesis that black
holes with the above dilaton couplings arise in string theory as  bound
states of  fundamental $a=\sqrt{3}$ states with zero binding energy. We also
generalize the result to states where the action does not admit  a single
scalar truncation and  show that a wide class of dyonic black holes in
toroidally compactified string theory can be viewed as  bound states of
fundamental $a=\srt$ black holes.


{\vfill
\leftline{December 1995}
\vskip 10pt
\footnoterule
{\footnotesize
 \hoch{\dagger} e-mail: joachim@tam2000.tamu.edu}

\newpage
\section{Introduction}
\la{intro}

The role of black holes in string theory is a  topic of active research
\cite{Rahmfeld1, Khurinew, Senblacktorus, Senblackholes, Greenecond, Cvetic,
Ramziscatter,
Liutrial, Callanblack, Harveyblack, Kalloshforce, Susskind}. In one of the
earliest papers discussing the  relation between extremal black holes and
massive string states \cite{Rahmfeld1} the observation was made that the
charge and mass quantum numbers of special extremal black holes are
consistent with a bound state interpretation. Those solutions could each be
described by an effective action
\be
 I_a={1\over16\pi G}\int d^4x\sqrt{-g}\left[R-\frac{1}{2}(\partial\phi)^2
-{1\over4}e^{-a\phi}F^2\right],
\la{actiona}
\ee
where $a$ is the dilaton coupling parameter, $\phi$ is an effective dilaton,
and $F$ is an effective field strength. $\phi$ and $F$ are typically linear
combinations of a variety of fields in the underlying theory. The couplings
consistent with supersymmetry are $a=\sqrt{3},1,\frac{1}{\sqrt{3}},0$
\cite{Rahmfeld1, Gibbons, Liutrial, Popemax, Khurisusy}. Incidentally,  the
extremal black hole solutions with precisely these values of $a$ have
regular null surface \cite{Hull}.

The conjecture was  that two, three or four fundamental $a=\sqrt{3}$ black
holes (with appropriate charges) can combine with neutral binding energy to
create  $a=1,\oosrt$ or $a=0$ states. For example, the extreme
Reissner-Nordstr\"om black hole is interpreted as a bound state of four
Kaluza-Klein black holes! Our goal is to prove this hypothesis and see
whether we can extend it to a wider class of solutions.

The starting  point of every discussion is the bosonic action of heterotic
string theory toroidally compactified to $D=4$ and broken to $U(1)^{28}$. In
canonical coordinates it is given by
\begin{equation}
I_H={1\over16\pi G}\int d^4x\sqrt{-g}\left[R-\frac{1}{2}(\partial\eta)^2
-{e^{-2\eta}\over12}H^2+{1\over8}\Tr(\partial  M
L\partial M L)-{1\over4}e^{-\eta}F{}^T
( L M L)F\right],
\la{full}
\end{equation}
where $L$ is the metric of $O(6,22)$. $M=M^T\in O(6,22)/O(6)\times O(22)$
parametrizes the scalars in the sigma model, the 28 $F_{\mu\nu}$'s are the
$U(1)$ fields strengths, and $\eta$ is the four dimensional dilaton. We use
the conventions of \cite{Senreview, Rahmfeld1}. Let us work on a special
point in the moduli space and set the asymptotic value of $M$ to
$M^{(0)}=I$. All other backgrounds are equivalent to this one by an
$O(6,22)$ T-duality transformation. T-duality also allows further
simplifications: for the solutions we are interested in, charged black
holes, we can use $O(6)\times O(22)$ transformations to truncate all field
strengths but four, $F_1,  \ F_2, \ F_3$ and $F_4$. The first two are
Kaluza-Klein fields, the other two are winding modes. It is noteworthy that
T-duality ensures that the only relevant degrees of freedom arise from the
compactification from six to four dimensions on a torus.

With axion-like fields set to zero we arrive at \bea I_t&={1\over16\pi
G}\int d^4x\sqrt{-g}\biggl
\{&\hspace{-0.4truecm}R-\frac{1}{2}\left[(\partial\eta)^2
+(\partial\s)^2+(\partial\rho)^2\right]-\nn\\ & & \hspace*{-0.5truecm}
-\frac{e^{-\eta}}{4}\left[e^{-\s-\r}F_1^2-e^{-\s+\r}F_2^2
-e^{\s+\r}F_3^2 -e^{\s-\r}F_4^2\right] \biggr\}. \la{action}
\eea
$e^{-\sigma}$ and $e^{-\rho}$ belong to  the K\"ahler form and complex
structure of the torus. This action was thoroughly analyzed in
\cite{Liutrial}. Especially, the triality of the three $SL(2)$  duality
groups was emphasized.

The extremal black holes we have in mind to illustrate the point, namely
those that allow for a description by an action of type (\ref{actiona}) with
$a=\sqrt{3},1,\oosrt,0$, have the following electric/magnetic charge quantum
numbers and masses \cite{Rahmfeld1, Liutrial}:

\bigskip

\noindent
(i)  $a=\srt$: $q=(1,0,0,0)$, $p=(0,0,0,0)$, $m=1,$

\bigskip

\noindent
(ii) $a=1$: $q=(1,0,1,0)$, $p=(0,0,0,0)$, $m=2 ,$

\bigskip

\noindent
(iii) $a=\oosrt$: $q=(1,0,1,0)$, $p=(0,1,0,0)$, $m=3 ,$

\bigskip

\noindent
(iv) $a=0$: $q=(1,0,1,0)$, $p=(0,1,0,1)$, $m=4,$

\medskip
where $q_A$ and $p_A$ are the electric and magnetic charges of field
strength $F_A$. These charges  and masses are certainly consistent with the
bound state interpretation. However, to prove the hypothesis we have to show
that there is a no-force condition between the relevant fundamental black
holes, since a  look at the masses shows that the binding energy vanishes.
The best way to guarantee vanishing total force  is to find a static
multi-black hole solution (with four single black holes) for arbitrary
distances between the constituents. This solution should have the correct
limits if one, two or three black holes are pushed out to infinity. Recent
papers discussed cases of that kind by making use of the chiral null model
\cite{Horowitz}. In \cite{Behrndt1, Behrndt2, Bekal} solutions were
constructed which are  relevant for the $a=1$ case. In this letter, we will
extend the idea to include dyonic charges to account for the other two
dilaton couplings. Our results are consistent with \cite{Holzhey} where it
was argued that only dilatonic black holes with $a>1$ can behave like
elementary particles.

Most black holes in the string spectrum cannot be described by a single
scalar action.  The natural question arises whether the bound state
interpretation can be generalized to these solutions as well. The answer is
yes, at least for a certain class, as will be discussed in section
\ref{general}.

\section{The Basic Multi-Black Hole Solution}
\la{multisol1}

Let us look for solutions of (\ref{action}) (one has to keep in mind
that axion-like fields are set to zero which imposes some constraints). 
The equations of motion of  are:
\bea
\nabla_\mu\left(e^{-\eta-\s-\r}
F_1^{\mu\nu}\right)& = & 0, \nn \\
\nabla_\mu \nabla^\mu \eta&=&
-\frac{1}{4}e^{-\eta}\left(e^{-\s-\r}F_1^2+
e^{\s+\r}F_3^2+e^{-\s+\r}F_2^2+e^{\s-\r}F_4^2\right),\nn  \\
R_{\mu\nu}& = & \frac{1}{2}\left(
   \partial _\mu \eta \partial_\nu \eta   +\partial _\mu \s\partial_\nu \s+
\partial _\mu\r \partial_\nu \r  \right)  + \nn \\
 & &\hspace*{-1.4truecm} +\frac{e^{-\eta-\s}}{2}\left[e^{-\r}\left(F_{1\mu\l}
  F_{1 \nu}^{\phantom{1 \mu}\l}-\frac{1}{4}g_{\mu\nu}F_1^2\right)
+e^{\r}\left(F_{2\mu\l}
  F_{2 \nu}^{\phantom{1 \mu}\l}-\frac{1}{4}g_{\mu\nu}F_2^2\right)\right]+
 \nn \\
& & \hspace*{-1.4truecm} +\frac{e^{-\eta+\s}}{2}\left[e^{\r}\left(F_{3\mu\l}
  F_{3 \nu}^{\phantom{1 \mu}\l}-\frac{1}{4}g_{\mu\nu}F_3^2\right)
+e^{-\r}\left(F_{4\mu\l}
  F_{4 \nu}^{\phantom{1 \mu}\l}-\frac{1}{4}g_{\mu\nu}F_4^2\right) \right]
\la{eom}
\eea
plus three additional Maxwell and two scalar equations.

To demonstrate the pattern, let us briefly review the  solutions
that can alternatively be described by (\ref{actiona}). With
\be
\Delta=(1+\frac{Q}{r})^{\frac{1}{2}}, \, \, r=\sqrt{x^2+y^2+z^2}.
\ee
the solutions are:

\medskip

\bea
{\rm{(i)}}\ a=\srt: \hspace{1.3truecm} ds^2 & = & -\D^{-1} dt^2 +\D
(dx^m dx^n \delta_{mn})\nn\\
e^{-\eta}&=&\D,  \, \, e^{-\s}=\D, \, \, e^{-\rho}=\D \la{asr3}
 \\
F_{1 tm}&=& \frac{Q x^m}{r^3 \D^4}\nn \\
\nn \\
\nn \\
{\rm{(ii)}}\ a=1: \hspace{1.5truecm} ds^2 & = & -\D^{-2} dt^2 +\D^2
(dx^m dx^n\delta_{mn})\nn\\
e^{-\eta}&=&\D^2,  \, \, e^{-\s}=1, \, \, e^{-\rho}=1\la{a1}  \\
F_{1 tm} &=& F_{3 tm} =\frac{Q x^m}{r^3 \D^4}\nn
\\
\nn \\
\nn \\
{\rm{(iii)}} \ a=\oosrt: \hspace{1truecm} ds^2 & = & -\D^{-3} dt^2 +\D^3
(dx^m dx^n \delta_{mn})\nn\\
e^{-\eta}&=&\D,  \, \, e^{-\s}=\D^{-1}, \, \, e^{-\rho}=\D \la{aoosrt}
 \\
F_{1 tm} &=& F_{3 tm} = \tilde{F}_{2 tm} =\frac{ Q x^m}{r^3 \D^4}\nn\\
\nn \\
\nn \\
{\rm{(iv)}} \ a=0: \hspace{1.4truecm} ds^2 & = & -\D^{-4} dt^2 +\D^4
(dx^m dx^n\delta_{mn})\nn\\
e^{-\eta}&=&1,  \, \, e^{-\s}=1, \, \, e^{-\rho}=1 \la{a0} \\
F_{1 tm} &=& F_{3 tm} = \tilde{F}_{2 tm} =\tilde{F}_{4 tm} =
\frac{ Q x^m}{r^3 \D^4}\nn
\eea
with $\tilde{F}_{2/4}=e^{-\eta\pm (-\s+\r)}F^*_{2/4}$, and where $^*$
denotes the Hodge dual. One should add that the solutions of type $a=\srt$
with other gauge fields (and possibly magnetic charges) are obtained from
(\ref{asr3}) by the obvious changes in $F$ and in the scalar fields, as
indicated  by (\ref{action}). $Q$ is understood to be quantized, since the
electric and magnetic charge vectors of string states belong to an even
self-dual lattice \cite{SchwarzSen, Senreview}.

The behavior of the metric is noteworthy: $g_{tt}=-\D^n$ and $g_{mm}=\D^n$
where $n$ is the number of conjectured constituents. Also, the scalar fields
seem to behave rather linearly in the contributions of the individual black
holes. This suggests a very simple solution for the four-black hole
configuration. It seems that the scalar and Maxwell fields should just be
added and the metric components multiplied.  For the case of  two extremal
$a=\sqrt{3}$ black holes, electrically charged under $F_1$ and $F_3$,
(\ref{a1}) coincides with a limit of the chiral null model, as discussed in
\cite{Behrndt1, Behrndt2}.

Let four black holes of type $a=\srt$ and electric charges in $F_1,
\tilde{F}_2, F_3, \tilde{F}_4$, respectively, be placed at $y^m_{A}$ with
$A=1,2,3,4$. The straightforward generalizations of $\Delta$ and $r$ are
\be \D_A=(1+\frac{Q}{r_A})^{\oh}, \ \, r_A=\sqrt{(x^1-y^1_A)^2+
(x^2-y^2_A)^2+(x^3-y^3_A)^2}.\la{single}
\ee
It is useful to  introduce shifted coordinates
$x^m_{A}=x^m-y^m_{A}$.
The solution of  (\ref{eom}) is
\bea ds^2 & = & -(\D_1 \D_2  \D_3  \D_4)^{-1}dt^2 +
 (\D_1 \D_2  \D_3  \D_4) (dx^m dx^n \delta_{mn}),\nn\\
e^{-\eta}&=&\frac{\D_1\D_3}{\D_2\D_4}, \, \,
e^{-\s}=\frac{\D_1\D_4}{\D_2\D_3}, \, \,
e^{-\r}=\frac{\D_1\D_2}{\D_3\D_4},\la{multi}
 \, \,
\\
F_{1/3 \ tm} &=&\frac{ Q x^m_{1/3}}{r_{1/3}^3 \D_{1/3}^4}, \nn\\
\tilde{F}_{2/4 \ tm} &=&\frac{ Q x^m_{2/4}}{r_{2/4}^3 \D_{2/4}^4}. \nn
\eea
It interpolates between the extremal black holes we are
targeting. In the  limit of $n$ constituents at the origin and $4-n$
at infinity,
we recover the standard $a=\sqrt{\frac{4}{n}-1}$ solutions.
Note that in this
simple case $a$ reveals a new interpretation:  $a^2$ is
the ratio of black holes at infinity and black holes at the origin.

The supersymmetry
breaking is determined by the central charges. Since mass and
charge are evaluated in the
asymptotic  region, the multi-black hole solution preserves
as many supersymmetries
as the appropriate single dilatonic black hole.

Another feature is also interesting: from the supersymmetry point of view,
the simple bound state interpretation seems to break down after we reach
$n=4$ constituents with distinct charges. This finds its correspondence in
the solutions of (\ref{actiona}), the power of $\D^{-1}$ in the metric is 
limited by $4$, since
\be g_{tt}=-\D^{-\frac{4}{1+a^2}}!\ee

It seems very surprising to find such a simple multi-black hole solution,
since the Einstein equations are non-linear. After all, with the ansatz
\be
ds^2=-f^{-1}dt^2+f  dx^m dx^n \delta_{mn},
\ee
the non-vanishing components of the Ricci tensor become
\bea
R_{tt}&=&-\frac{1}{2 f^2}\S_k \partial_k \partial_k \ln f \nn \\
R_{ii}&=&-\frac{1}{2}\S_k\partial_k \partial_k \ln f -\frac{1}{2}
\partial_i \ln f \partial_i \ln f  \\
R_{ij}&=& -\frac{1}{2}\partial_i \ln f \partial_j \ln f, \nn
\eea
which is partially non-linear in the contributions of the individual
constituents. Also, the scalar terms on the right hand side of the Einstein
equations are  non-linear. However, for the solution (\ref{multi}) the
non-linearities on both sides cancel! Essentially, this can be seen by
noting that in
\be
 (\ln f)^2+\eta^2+\s^2+\r^2=4\left[(\ln\D_1)^2+(\ln\D_2)^2+
(\ln\D_3)^2+(\ln\D_4)^2\right]
\ee
the non-linearities on the left-hand side vanish. This cancellation
translates directly to the Einstein equations.

Since (\ref{multi}) is static it implies a no-force condition between all
four fundamental $a=\srt$ extremal  black holes. One might wonder what
cancels the gravitational attraction, after all the charges do not provide
the repulsion. It turns out that the scalars (or some of them) are repulsive
for our type of solution. This behavior is  analogous to the case considered
recently in \cite{Kalloshforce}.

The new  multi-black hole solution provides us with the proof for the bound
state interpretation of dilatonic black holes with the standard couplings.
However, we can push the idea even further, which is what we will do in the
next chapter.

\section{More General Dyonic Black Holes As Bound States}
\la{general}
It is well-known that, in general, black hole solutions in string theory
cannot be described by an effective single scalar, single gauge field action
\cite{Rahmfeld1}. The procedure outlined above shows that with ansatz
(\ref{multi}) the equations decompose into independent equations of motion
for states charged under a single gauge field. Therefore, we can use the
familiar  multi-monopole ansatz \cite{Ramzi, Harveymulti}, of which
(\ref{single}) was a special case, for each charge sector. In the most
general case, each individual $\Delta$  becomes
\be
\Delta=(1+\S_l\frac{Q_l}{|\vec{r}-\vec{r}_l|})^{\frac{1}{2}}
\la{multimo}
\ee
where $Q_l$ and  $\vec{r}_l$ denote the charges and positions of the black
holes with common gauge field. The gauge field itself is modified in the
well-known way. Overall, we obtain a static multi-black hole solution of
mixed type: a combination of linear superpositions of identical black holes
and a product combination of black hole sectors with different charges.

The interesting aspect of this more general ansatz is that it extends
the bound state interpretation  to solutions with arbitrary (quantized)
left- and right-handed charges. For simplicity, let us consider an example
with black holes electrically charged under $F_1$ or $F_3$ and take the limit
in which all black holes are located at the origin. The multi-black hole
solution then reduces to:

\bea
ds^2&=&-\left(1+\frac{Q_1}{r}\right)^
{-\frac{1}{2}} \left(1+\frac{Q_3}{r}\right)^{-\frac{1}{2}}
dt^2+\left(1+\frac{Q_1}{r}\right)^
{\frac{1}{2}} \left(1+\frac{Q_3}{r}\right)^{\frac{1}{2}}
 dx^m dx^n \delta_{mn},   \nn \\
e^{-\eta} &=& \left(1+\frac{Q_1}{r}\right)^
{\frac{1}{2}} \left(1+\frac{Q_3}{r}\right)^{\frac{1}{2}},\nn \\
e^{-\sigma}&=&e^{-\rho}=\left(\frac{1+\frac{Q_1}{r}}
{1+\frac{Q_3}{r}}\right)^{\frac{1}{2}}, \la{generalsol}\\
F_{1 \ tm}&=&\frac{ Q_1x^m}{r^3 \left(1+\frac{Q_1}{r}\right)^{2}  },  \nn\\
F_{3 \ tm}&=&\frac{ Q_3x^m}{r^3 \left(1+\frac{Q_3}{r}\right)^{2}  },  \nn
\eea
where $Q_1$ and $Q_3$ are the total charges of all black holes charged under
$F_1$ and $F_3$, respectively.
With
\bea
Q_R&=&\frac{1}{2}(Q_1+Q_3) \nn \\
Q_L&=&\frac{1}{2}(Q_1-Q_3)
\eea
it is clear that (\ref{generalsol}) can have arbitrary half-integer left-
and right-handed electric charges. A simple calculation reveals that
(\ref{generalsol}) agrees with Sen's  extreme non-rotating black hole
solutions \cite{Senblacktorus} for states electrically charged under $F_1$
and $F_3$.

The above example can be easily generalized to include magnetic charges, in
which case the multi-monopole ansatz (\ref{multimo}) applies to
$\D_2, \D_4, \tilde{F}_2$ and $\tilde{F}_4$ also. Again, the right- and
left-handed magnetic charges can take any half-integer value.

Overall, we find that the bound state interpretation can be extended
to dyonic states with charge and mass quantum numbers
\be
   q=(Q_1,0,Q_3,0),\ \  p=(0,Q_2,0,Q_4),\ \  m=Q_1+Q_2+Q_3+Q_4,
\la{dyon}
\ee
which, in general, do not admit  a single scalar truncation. Note  that the
mass/charge relation in (\ref{dyon}) does not agree with the Schwarz-Sen
mass formula \cite{SchwarzSen},
since generically,  dyonic black holes break more than one half of the
supersymmetries. However, $m$  is perfectly consistent with
\cite{Cvetic,Liutrial}.

\section{Conclusion}
In this letter, we discussed an extended multi-monopole solution which
established a no-force condition between four $a=\srt$ black holes, with
electric/magnetic charges in different gauge fields. The gravitational
attraction is balanced by scalar repulsion. This proved that black holes of
dilaton coupling $a=1,\oosrt,0$ can be viewed as  bound states of two, three
and four $a=\srt$ black holes with zero binding energy, as conjectured in
\cite{Rahmfeld1}. The new interpretation of these dilatonic black holes
explains very nicely the behavior of the metric. So far, there was no
physical understanding of the relative powers of $\D$ in $g_{tt}$ and
$g_{mm}$ of the extremal solutions with the discussed couplings. The
multi-black hole solution fills this gap, it  gives a  physical
interpretation of  the ``quantization'' of powers for the crucial $a$
values. Also, the coupling $a$ becomes a new physical meaning: $a^2$ is the
ratio of black holes at infinity and black holes at the origin.

It turned out that
the bound state interpretation  could be generalized  to a wide class of
dyonic black holes in string theory as well.

\bigskip

\bigskip

\bigskip

\noindent
{\large \bf Acknowledgments:}

\medskip

It is a pleasure to thank Michael Duff, Ramzi Khuri and Sudipta Mukherji
for inspiring conversations.

\bigskip

\bigskip

\bigskip

{\large \bf Note added:}

\medskip

After completion of this work we became aware of the work by
{Cveti\v c} and Tseytlin \cite{Cvettseyt}. They discuss extreme dyonic
single-centered  black hole solutions with the same charges as those in
the present work, also characterized by four harmonic functions.

\bibliographystyle{preprint}



\end{document}